%% file: template.tex
\documentclass[journal, onecolumn]{IEEEtran}
\usepackage[nolist]{acronym}
\usepackage{graphicx}
\usepackage{caption}
\usepackage{subcaption}

\begin{document}

\title{Experimental Demonstration of RAN Functional Split over virtual PON Transport Network}

\author{\IEEEauthorblockN{Line Larsen$^1$, Frank Slyne$^2$, Ghizlane Mountaser$^3$, Marco Ruffini$^2$, Toktam Mahmoodi$^3$}
	\IEEEauthorblockA{\\(1) Department of Photonics Engineering, Technical University of Denmark, Denmark \\ (2) CONNECT research centre, School of Computer Science and Statistics, Trinity College Dublin, Ireland \\ (3) Centre for Telecommunications Research, Department of Engineering, King's College London }
	}

\maketitle







\begin{abstract} 
Cloud-Radio Access Networks (Cloud-RANs) are separating the mobile network’s base station functions into three units, the connection between the two of them is referred to as the fronthaul network. This work demonstrates the transmission of user data transport blocks between the distributed Medium Access Control (MAC) layer and local Physical (PHY) layer in the radio unit over a Passive Optical Network (PON). PON networks provide benefits in terms of economy and flexibility when used for Cloud-RAN fronthaul transport. However, the PON upstream scheduling can introduce additional latency that might not satisfy the requirements imposed by Cloud-RAN functional split.
In this work we demonstrate how our virtual \ac{DBA} concept can be used to effectively communicate with the mobile \ac{LTE} scheduler, adopting the well known cooperative \ac{DBA}  mechanism, to reduce the PON latency to satisfactory values. 
Thus, our results show the feasibility of using PON technology as transport medium of the fronthaul for the MAC/PHY functional split, in a fully virtualised environment. Further background traffic is added, so that measurements show a more realistic scenario. The obtained round trip times indicates that using PON at fronthaul might be limited to the distance of $11$~km for a synchronised scenario, or no compliance for an non-synchronised scenario.
\end{abstract}


\input{acronym.tex}
\section{Introduction}
Cloud-Radio Access Network (Cloud-RAN) is a promising technology that is able to scale and enhance network efficiency to support new features over the next several years by means of centralisation. In Cloud-RAN architecture, baseband functions are centrally deployed providing several benefits in terms of offering high level of cooperation between base stations and facilitating RAN sharing. These aspects allow dynamic reconﬁguration of resources enabling to fulfil diverse requirements of various vertical industrial applications.
Cloud-RAN significantly benefits from the emerging technologies such as softwarisation and virtualisation. Softwarisation/virtualisation within Cloud-RAN allows adaptive allocation of RAN functions between components of Cloud-RAN introducing the so-called flexible functional split. Flexible functional split typically offer flexibility and adaptability in the Cloud-RAN, thus, enabling to deliver services with diverse requirements.
In NGMN \cite{NGMN2018}, Cloud-RAN is composed of Central Unit (CU), Distributed Unit (DU) and \ac{RU}. The communication link between CU and DU is called midhaul, and between DU and \ac{RU} is called fronthaul. NGMN proposed different allocations of the baseband functions between the three components. In particular, it deﬁned a high layer split (HLS) and a low layer split (LLS), each of them have different communication performance requirements. In this paper, we consider LLS, specifically Medium Access Control/Physical layer (MAC/PHY) split.

\begin{figure}[t]
\centering
\includegraphics[width=.8\linewidth]{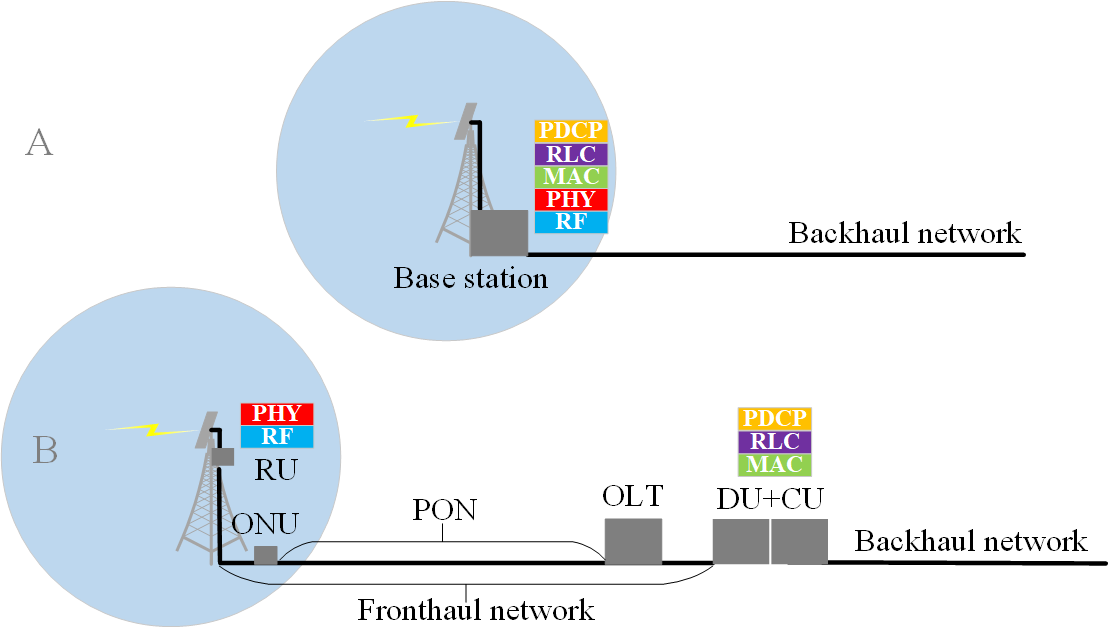}
\caption{(A) Traditional base station site. (B)  Cloud-RAN with PON fronthaul including the ONU and OLT. The MAC/PHY functional split is divided out on \ac{RU} and DU+CU as per NGMN central RAN LLS \cite{NGMN2018}. }
\label{fig:Fig1l}
\end{figure}

\par Different solutions exist for LLS transport interface (fronthaul). 
The benefits of the fiber fronthaul network, if available, are multiple folds and includes a longer transmission range, a low loss transmission medium and the opportunity of very high transmission capacity. Given the high cost of fiber deployments, research is looking into the use of fibers that have already been installed for other purposes, and investigates how the fronthaul network can co-exist with other demanding applications. A network technology widely deployed for fiber to the home (FTTH) solutions is Passive Optical Network (PON)  \cite{Lee2013}. 
\par The PON network consists of one or multiple \acp{ONU} connected to one \ac{OLT}. Figure 1A illustrates the traditional base station ( the Distributed RAN - D-RAN) and Figure 1B illustrates Cloud-RAN with MAC/PHY functional split transported over a PON fronthaul. In a PON network, multiple technologies can be multiplexed leaving the opportunity of connecting multiple \acp{RU} with different functional splits and both constant and variable fronthaul bit rate \cite{Nomura2017}. Nevertheless, PON has one problem when acting as a fronthaul network; the mechanism that  guarantees dynamic and exact scheduling of upstream resources between the \ac{OLT} and every \ac{ONU} is based on \ac{DBA}. \ac{DBA} ensures high bandwidth utilization but requires the \ac{ONU} to request bandwidth at the \ac{OLT} before each upstream transmission \cite{Nomura2017}. This work aims at overcoming this challenge by presenting an optimised DBA solution, referred to as the cooperative DBA, whereby bandwidth allocation in PON is requested when upstream scheduling information is sent to the UE. With this solution, we demonstrate the feasibility of transporting the MAC/PHY split over an \ac{XGS-PON} fronthaul network  using a hardware-based test-bed. 

This paper is organized as follows: Section \ref{sec:state} provides an overview of the status of current work in the area of PON fronthauling. Section \ref{sec:fronthaul} introduces the XGS-PON fronthaul. Section \ref{sec:testbed} presents the test-bed used for measurements in this work. Section \ref{sec:feasibility} propounds the first round of feasibility studies. Section \ref{sec:comp} propounds extended studies where background traffic is included and compares the cooperative DBA scenario to the original non-synchronised scenario. Finally section \ref{sec:conc} concludes the paper.

\section{State of the art}
\label{sec:state}
Recent studies have looked into the idea of using PON as the fronthaul network for Cloud-RAN. One of the first works considering PON fronthaul is found in \cite{Tashiro2014}, which states the economical benefits of using PON fronthaul for small cells. The work in \cite{Tashiro2014} introduces the concept of using mobile upstream scheduling information for reduction of Round Trip Time (RTT) delay. The work in \cite{Tashiro2014} considers a simple functional split with only the RF part in the \ac{RU}, and the PHY and upper transmission processing in the CU. The concept of this current work relies on the work in \cite{Nomura2017} where the authors present an optical-mobile cooperation interface converting mobile upstream scheduling information to transmit request information for the PON. The idea is to transmit the request information to the \ac{OLT} directly from the CU. In \cite{Nomura2017} the actual functional split used for the measurements is not specified, but the paper states the benefits of a packet switched fronthaul network. The work in \cite{Nomura2017} extends the conversion process to handle multiple input frames and considers multiple \acp{RU}. In \cite{Zhou2018}, the authors demonstrate a simulated transmission of PHY functional split over PON. In \cite{Das2019} the PON fronthaul is used as a variable rate fronthaul for the functional split with the lowest number of functions in the \ac{RU}, having a constant fronthaul bit rate. Using a software defined network controller, the cell bandwidth and consequently the fronthaul bit rate is dynamically changed. In \cite{Alvarez2018} the variable-rate PON fronthaul is experimentally demonstrated.\\ Some of the authors of this current work have in \cite{1Mountaser2017} experimented with transport of transport blocks between the PHY in the \ac{RU} and the MAC in the CU over an Ethernet fronthaul network. The same test-bed transporting MAC/PHY split transport blocks over Ethernet is further investigated in \cite{2Mountaser2017} for \ac{URLLC} scenarios. The work in \cite{Ou2016} presents a model deriving the achievable latency for a PON fronthaul and the maximum number of connected \acp{RU}. This model forms the basis of the latency calculations in this work. For a comprehensive overview of all functional splits is referred to \cite{Larsen2019}. Regarding the MAC/PHY functional split, NGMN defines this interface location more specifically as between the MAC and the Forward Error Correction (FEC) in the PHY \cite{NGMN2015}. When using the MAC/PHY functional split, transport blocks are transferred over the fronthaul network \cite{Larsen2019}. These transport blocks have been through the physical layer processing which means the upstream signal is, among other processes, de-mapped, demodulated an de-scrambled before transmission \cite{Larsen2019}. The MAC/PHY functional split is defined by NGMN as one of the converging options for LLS \cite{NGMN2019}. Multiple options for LLS are being developed in parallel providing different split options to support different use cases \cite{NGMN2019}. The MAC/PHY functional split is beneficial in use cases requiring a low fronthaul bit rate and a centralised scheduler coordinating the available resources \cite{Larsen2019}. The MAC/PHY functional split has time critical functions located in the DU+CU and this puts stringent requirements on the fronthaul latency \cite{Larsen2019}.  NGMN \cite{NGMN2015} defines a maximum one-way latency of 250 us, which is at the protocol stack level. 

\section{The \ac{XGS-PON} Fronthaul}
\label{sec:fronthaul}
The type of PON we use in this work is \ac{XGS-PON} \cite{ITU-T}. In a PON, the \ac{OLT} handles the traffic management and in the downstream direction, the \ac{OLT} transmits data to the \ac{ONU} while being aware of the quality of service. In the upstream direction, the user sends data to the \ac{ONU} and the \ac{ONU} stores the data in a queue until it can be forwarded. When the \ac{ONU} receives the user data it sends a DRBu message to the \ac{OLT} requesting bandwidth allocation for the specific queue size. The \ac{OLT} will then reply with a \ac{BMap} message containing the start time and grant size. The upstream data is thereby delayed while it waits in the ONU’s queue. This delay is dependent on the transmission delay between the \ac{ONU} and OLT, and the time it takes for the \ac{OLT} to calculate the bandwidth assignment for the grant.

\begin{figure*} [t]
	\centering
	\begin{subfigure}{0.45\textwidth}
	{\includegraphics[width=.9\linewidth]{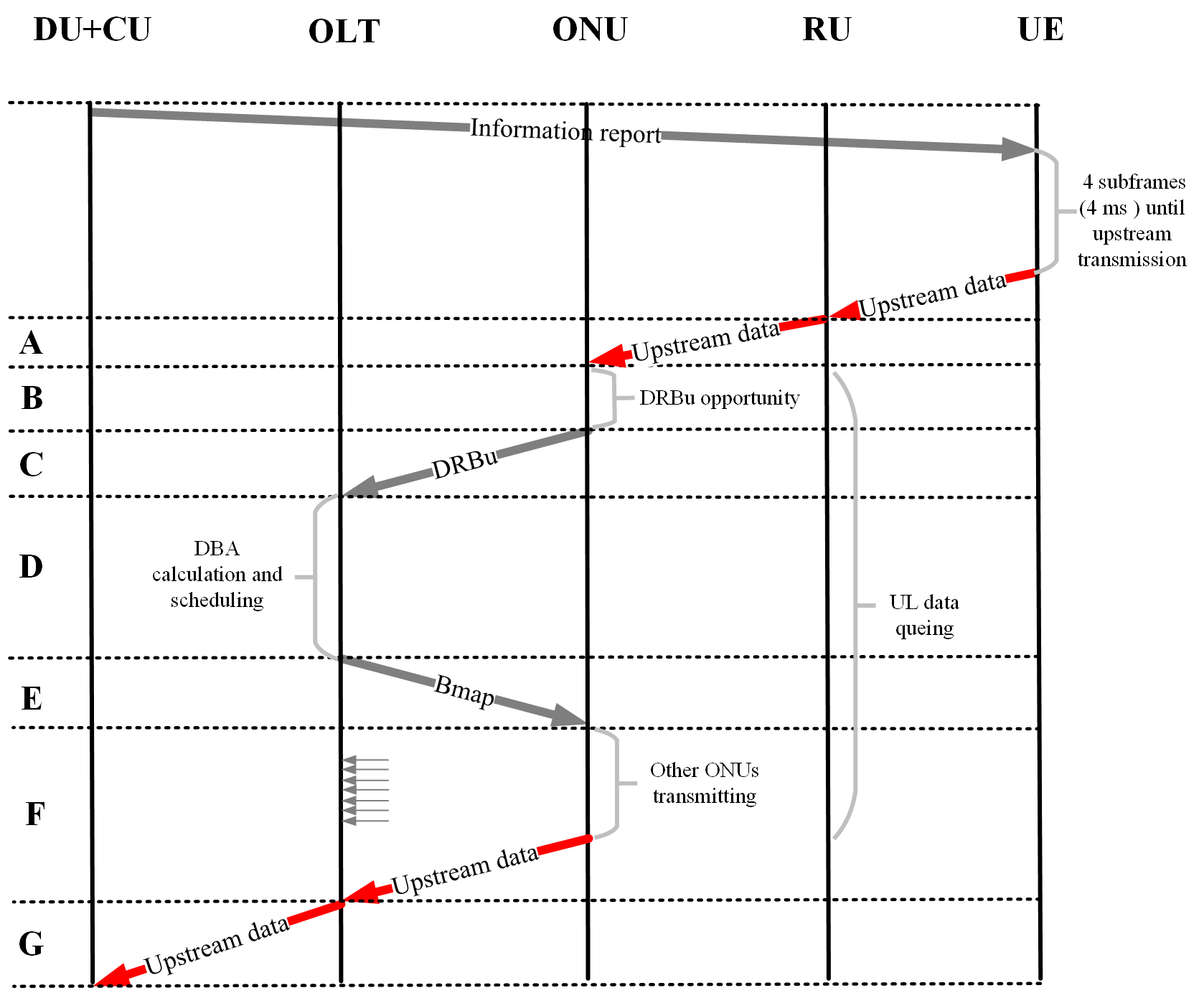}}
	\caption{}
	\label{fig:Fig2l}
	\end{subfigure}
	\begin{subfigure}{0.45\textwidth}
	{\includegraphics[width=0.9\linewidth]{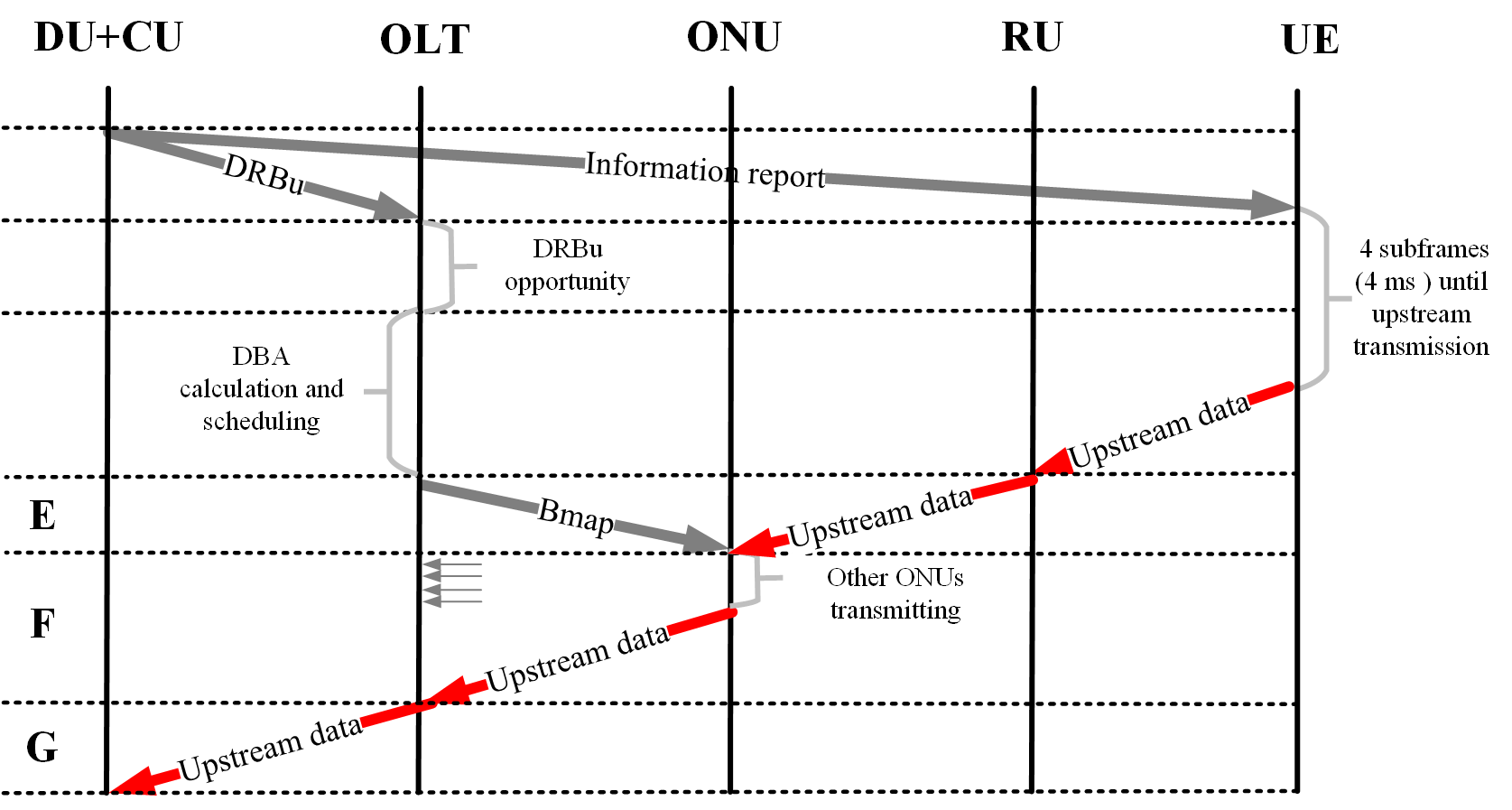}}
	\caption{}
	\label{fig:Fig3l}
	\end{subfigure}
	\caption{(a) The un-synchronised scenario showing the message flow between DU+CU and \ac{RU} in Cloud-RAN when introducing a PON in-between. In the figure the latency parameters are grouped into sections. The figure shows the large latency induced when the upstream data is queuing in the \ac{ONU} buffer. \\ (b) The cooperative DBA solution of PON fronthaul. In the figure the latency parameters are grouped into sections, showing that compared to figure 2, then this scenario have much less latency intervals.}
	\label{fig:alg_diff_high_800m}
\end{figure*}

 

\begin{figure*}[t]
\centering
\includegraphics[width=.75\textwidth]{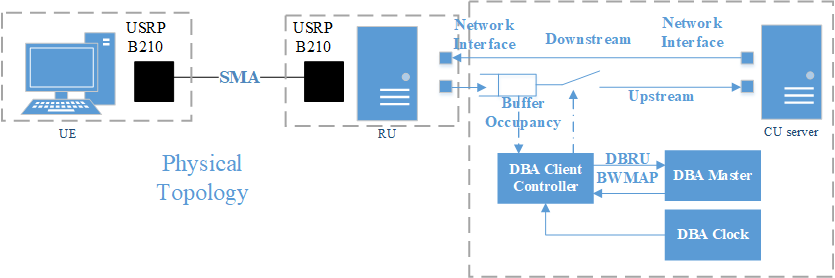}
\caption{The experimental setup showing the physical connections.}
\label{fig:Fig4l}
\end{figure*}

\begin{figure*}[t]
\centering
\includegraphics[width=.75\textwidth]{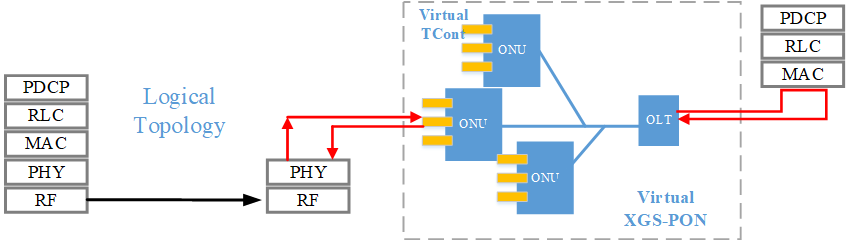}
\caption{The experimental setup showing the different functions in the protocol stack illustrated. A red line illustrates the RTT, measured from the upstream message leaves the \ac{RU} until an ACK is received.}
\label{fig:Fig5l}
\end{figure*}

The delay from waiting in the \ac{ONU} queue is normally not a problem when considering FTTH solutions suited to domestic and business applications, for instance Video, Voice and Data traffic. However the timing constraints to carry mobile networks, such as \ac{LTE} and 5G are much more time critical. In \cite{NGMN2015} NGMN is recommending a maximum one-way latency of 250 us  which we consider for our analysis. 
This gives a maximum RTT of 500 us for the PON fronthaul transmission.To reduce latency on the fronthaul for upstream transmission, this work makes use of the concept from \cite{Nomura2017}, called Cooperative DBA, recently standardised in \cite{ITU-CONV}, where the DU+CU, instead of the ONU, sends the DRBu message to the OLT. This solution is possible because in mobile networks an information report is created in the upper layers of the protocol stack, informing the UE of how many Transport Blocks (TBs) are allocated to the specific transmission. This cooperative DBA scenario is illustrated in figure \ref{fig:Fig3l}. The un-synchronised scenario, where there is no cooperation between PON an LTE, is illustrated in \ref{fig:Fig2l}. In this work, the data from the information report is converted to a \ac{DBRu} message inside the DU+CU, and sent from the DU+CU to the OLT, straight after the information report is sent. This way, the \ac{ONU} will receive the \ac{BMap} message earlier in the process and this will reduce the time spend by the user data in the \ac{ONU} queue. Also, when the UE receives the information report, it triggers an uplink transmission in 4 sub-frames later.

The delay budget for the un-synchronised scenario is divided into six intervals, illustrated in figure \ref{fig:Fig2l}:
\begin{enumerate}
	\item Send upstream data from \ac{RU} to ONU. In this interval, the upstream data is transmitted over the interface between the \ac{RU} and ONU. This interval includes time for transmission and propagation of a packet.
	\item Wait for DRBu upstream opportunity delay, because the DBRu is piggy backed into the current data transmission opportunity. The maximum delay is one frame.
    \item Time to send \ac{DBRu} from \ac{ONU} to OLT. In this interval the \ac{ONU} is transmitting a \ac{DBRu} message to the \ac{OLT} based on the queue size within the ONU. This interval includes time for transmission and propagation of a packet.
    \item Time for \ac{DBA} calculation and scheduling. This is the time the \ac{OLT} spends calculating the \ac{DBA}  for the specific ONU.
    \item Time to send \ac{BMap} from \ac{OLT} to ONU. In this interval, the \ac{BMap} is transmitted over the PON interface between the \ac{OLT} and ONU. This interval includes time for transmission and propagation of the \ac{BMap}.
    \item Time to wait for other \ac{ONU} transmissions and transmit upstream data for OLT. the \ac{ONU} tak delay. When the \ac{ONU} receives the \ac{BMap} it must wait until the allocated time. While the \ac{ONU} waits, other ONUs are transmitting to the OLT. 
    \item Send upstream data from \ac{OLT} to DU+CU. In this interval, the upstream data is transmitted over the interface between the \ac{OLT} and DU+CU. This interval includes time for transmission and propagation of a packet. The latency within the communication stack on the OLT, DU and CU are separate.
\end{enumerate}
The total latency $L_{un-synchronised}$ for the un-synchronised upstream transmission includes all intervals, illustrated in equation 1:
\begin{equation}
L_{un-synchronised} = \sum_{A}^{G} Latency
\label{eq:Elab}
\end{equation}
The total latency $L_{cooperative DBA}$ for the cooperative DBA upstream transmission includes only intervals E, F and G, illustrated in equation 2:
\begin{equation}
L_{cooperative DBA} = E + F + G
\label{eq:elabel}
\end{equation}

\section{Test-bed setup}
\label{sec:testbed}
Figure \ref{fig:Fig1l}B represents a logical overview of the test-bed setup. The \ac{RU} contains the radio frequency component and the physical layer.  In our test-bed implementation, the whole PON fronthaul network is collocated with the DU+CU on the same server, as illustrated in figure \ref{fig:Fig4l}. The \ac{XGS-PON} is a virtualised system, in that the key functionality related to the data plane transmission is rendered in software. In the upstream, data is packetised and \ac{TDM} multiplexed in a transmission path scheduled by a \ac{DBA} scheduler. This means that only one \ac{ONU} can transmit at a time. In the downstream, data is instead transmitted over a continuous frame and is broadcast from the \ac{OLT} to each \ac{ONU}. Outwardly, from the perspective of the \ac{LTE} application, the virtual \ac{XGS-PON} appears as an operating system level network interface on the servers for the DU+CU and the RU.  Upstream traffic enters a buffer, which is gated by a \ac{DBA} client controller. The \ac{DBA} client controller reports the number of bytes in the buffer, also called the Buffer Occupancy, to the \ac{DBA} master. The \ac{DBA} master collects the Buffer Occupancy reports from the total number of \acp{T-CONT} and applies a \ac{DBA} algorithm to decide the allocation start and stop time for each \ac{T-CONT}. 
\par The \ac{XGS-PON} emulates the characteristics of the delay budget outlined in figure \ref{fig:Fig2l} for both un-synchronised and cooperative DBA scenarios. In order to make the scenarios more realistic, we apply background loading to the \ac{XGS-PON} in section \ref{sec:comp}. This has the  effect of  inducing  packet loss, jitter and latency on the foreground \ac{LTE} traffic of interest.  For our emulation, the upstream \ac{LTE} traffic of interest as well as other similar \ac{LTE} traffic is marked as high priority traffic (that is Traffic Class 4) and makes up to 60\% of the upstream traffic. Other lower priority traffic (Traffic Classes 1,2 and 3) makes up the remainder (40\%) of the upstream traffic. The authors in \cite{Ruff2020} analysed how different mixes of upstream traffic priorities affect each other. 

The physical setup is illustrated in figure \ref{fig:Fig4l}. The \ac{UE} is a \ac{LTE} UE based on Open Air Interface (OAI) \cite{Nikaein2014} software further described in \cite{1Mountaser2017}. The UE server is connected to an Universal Software Radio Peripheral (USRP) B210 via an USB3 connection. The USRP is connected via a SubMiniature version A (SMA) cable to another USRP connected to the \ac{RU} server. On the \ac{RU} server is all radio frequency and physical layer processing taking place. The \ac{RU} server is connected to the DU+CU server via an Ethernet cable. The DU+CU server contains both the virtual \ac{XGS-PON} network and the DU+CU. Entering the DU+CU server the Ethernet packets are sent to the \ac{ONU}. In the \ac{ONU}, the Ethernet packets are encapsulated in PON \acp{T-CONT} for transmission over the XGS-PON. First they are queued, waiting for the \ac{OLT} to transmit the \ac{BMap}. When this is received, the traffic containers are virtually transmitted to the \ac{OLT} where they are unpacked to Ethernet frames and sent to the DU+CU. In the DU+CU the MAC and further baseband processing takes place. The DU+CU also creates DRBu messages based on information reports.
Figure \ref{fig:Fig5l}  shows how the RTT is measured from the \ac{RU} to the CU and back. RTT is used because it is a more accurate means of measurement in this specific test setup, when compared to one-way latency. \\
In our test-bed configuration, we implement three scenarios. Firstly, we implement a scenario where the \ac{XGS-PON} operates unaware of the higher level \ac{LTE} application. In particular, the upstream \ac{XGS-PON} and \ac{LTE} MAC schedulers operate independently of each other, referred to as the un-synchronised scenario. 
Secondly, we implement the cooperative DBA scenario, where the DU+CU issues the information report or DRBu message to the \ac{DBA} manager, which preemptively allocates the bandwidth to the \ac{T-CONT} handling the upstream \ac{LTE} traffic. This is referred to as the cooperative DBA scenario.
Thirdly we introduce the near ideal scenario, where LTE traffic is prioritised to reduce latency as much as possible. In theory, the upstream latency can be reduced to just upstream fibre transmission plus  some  opportunity delay, of around 35 usec on a 5 km fibre. Firstly, when the UL data arrives at the ONU, it is scheduled to take the best available slot, without waiting for the next PON frame. Secondly, because the LTE scheduler information is made available in advance, the transmission of the (n)th UL data frame can be predicted by from the timing of the (n-1)th dbru indicator from the LTE mac layer. 
\begin{table}[t]
\centering
\caption{\bf Physical and Emulation Parameters}
\begin{tabular}{ccc}
\hline
Parameter & Value \\
\hline
Carrier Frequency & 2.68 GHz  \cite{1Mountaser2017} \\
System bandwidth & 5 MHz  \cite{1Mountaser2017} \\
Frame type & FDD \cite{1Mountaser2017} \\
\# Upstream Tx/Rx antennas & 1 / 1  \cite{1Mountaser2017} \\
Tx Gain & 100  \cite{1Mountaser2017} \\
Rx Gain & 80  \cite{1Mountaser2017} \\
\ac{XGS-PON} max transmission speed & 10 Gbps\\
Fibre transmission latency of PON & 5 uSec/km\\
Upstream Scheduling Delay, worst case (E) & 15 uSec \cite{Ruff2020}\\
\hline
\end{tabular}
  \label{tab:ta1}
\end{table}


\section{Feasibility studies}
\label{sec:feasibility}
\begin{figure*} [t]
	\centering
    \includegraphics[width=.45\linewidth]{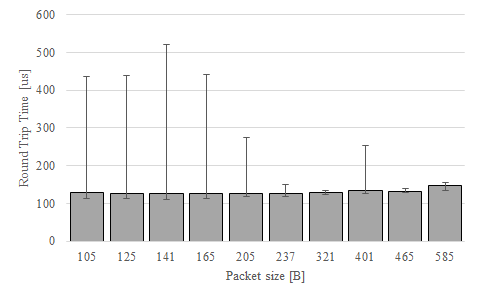}
    \caption{RTT latency obtained for different sizes of packets. The error bars shows the maximum and minimum RTT obtained.}
    \label{fig:Fig6l}
\end{figure*}

\begin{figure*} [t]
	\centering
	\begin{subfigure}{0.45\textwidth}
	{\includegraphics[width=.9\linewidth]{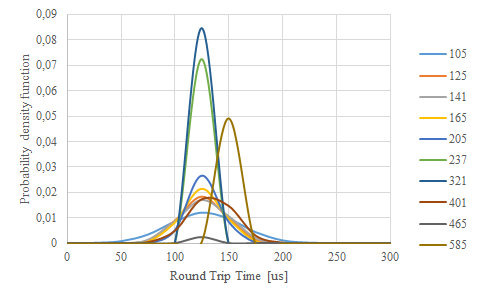}}
	\caption{}
	\label{fig:Fig7l}
	\end{subfigure}
	\begin{subfigure}{0.45\textwidth}
	{\includegraphics[width=0.9\linewidth]{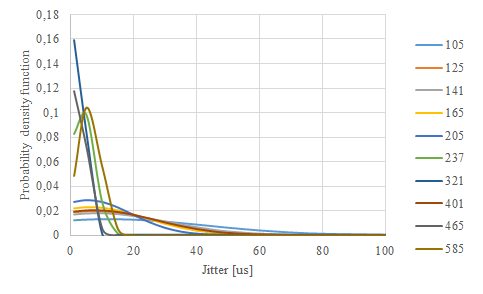}}
	\caption{}
	\label{fig:Fig10l}
	\end{subfigure}
	\caption{(a) Probability of latency for different sizes of packets. (b) Probability of jitter for different sizes of Ethernet packets }
\end{figure*}

First we demonstrate the feasibility of transporting the MAC/PHY functional split over the XGS-PON network using the cooperative DBA scenario. The physical and emulation parameters for the test setup are stated in table \ref{tab:ta1}. Figures \ref{fig:Fig6l} to \ref{fig:Fig10l} illustrate the feasibility of this goal, where the length of the XGS-PON network is one km. 


Figure \ref{fig:Fig6l} shows the RTT latency measured for different packet sizes between 105 B and 585 B. The tendency shows a slight increase in RTT latency when packet sizes are growing. In figure \ref{fig:Fig6l}, the average RTT is illustrated by columns and the lines shows the minimum and maximum RTTs. In general the smaller packets have larger fluctuations, but their average RTT is still low. The RTT measurements show how the average latency is compliant with the NGMN requirements stated in section \ref{sec:fronthaul} of 500 us. We believe this is caused by the variability on the UDP socket timing, which operates the functional split. This could be significantly reduced in the future by optimising the packet transfer process across functions. The maximum packet size measured is 585 B, which is due to upstream modulation restrictions in the OAI test-bed.


However, from Figure \ref{fig:Fig7l}, which  illustrates the RTT latency probability density function, we see that most transmissions are less than 150 us. The figure shows how a deviation occurs when the packet size is larger. Most of the different packet sizes have the highest probability for a RTT between 100 us and 150 us, which is still far below the NGMN threshold.


Figure \ref{fig:Fig10l}  illustrates the probability density distribution of the jitter, which is another limiting parameter for the MAC/PHY functional split. In this setup, which is a combination of Ethernet and virtual PON transmission, the data is transmitted in both Ethernet packets and virtual PON T-CONTs. This setup might induce some jitter due to synchronisation ad conversion between the different transport networks. 


The tendency shows how larger packet sizes have more probability for jitter, where the jitter in the smaller packets span over a longer period. Though figure \ref{fig:Fig10l} states that the jitter occurring is mainly below 20 us.

\section{Comparisons of different scenarios}
\label{sec:comp}
\begin{figure*} [t]
	\centering
	\begin{subfigure}{0.45\textwidth}
	{\includegraphics[width=.9\linewidth]{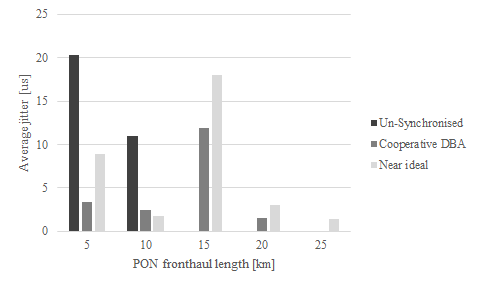}}
	\caption{}
	\label{fig:FigJitter}
	\end{subfigure}
	\begin{subfigure}{0.45\textwidth}
	{\includegraphics[width=0.9\linewidth]{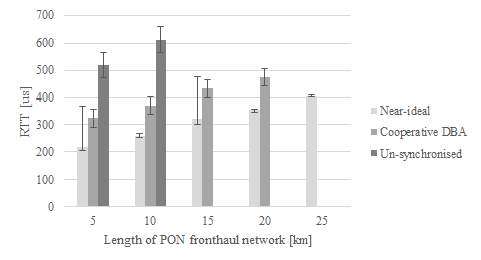}}
	\caption{}
	\label{fig:FigRTT_bar}
	\end{subfigure}
	\caption{(a)The average jitter for the near ideal, cooperative DBA and un-synchronised scenarios illustrated for different fronthaul lengths. (b) RTT of the near ideal, cooperative DBA and un-synchronised scenarios.}
\end{figure*}

 In this section the background traffic introduced in section \ref{sec:testbed} is included in the measurements in order to make the scenarios more realistic. The measurements were performed for different lengths of the XGS-PON fronthaul ranging from 5 km to 30 km. Results are illustrated in figures \ref{fig:FigJitter}, \ref{fig:FigRTT_bar}, respectively, for the jitter and RTT for increasing value of fronthaul distance. The figures illustrates the large difference between the three scenarios and how the RTT increases by the length of the fronthaul. The figure is only containing measurements for packet size 141 B.
 In the figures we can see that the un-synchronised scenario allows successful transmission up to 10 km distance, the cooperative DBA up to 20km and the near ideal scenario up to 25km.

\begin{figure} [t]
 \centering
  \includegraphics[width=.45\linewidth]{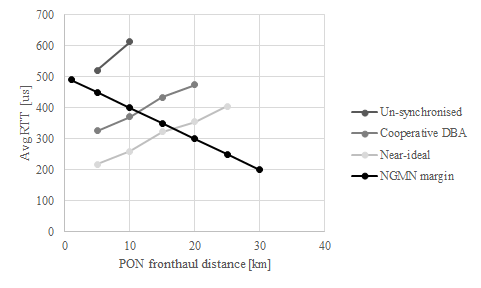}
  \caption{Margin to NGMN time limit of different PON fronthaul distances comparing the near ideal, cooperative DBA and un-synchronised average RTT scenarios to the NGMN margin times.}
  \label{fig:NGMN}
\end{figure}

While we found that transmissions could still work for latency up to 600 us, the NGMN \cite{NGMN2015} recommends a maximum one-way transport latency of 250 us, and the test-bed only measures timing in RTT, we assume the RTT threshold to be 500 us. The NGMN transport latency includes: UE processing time, eNodeB processing time and propagation time \cite{NGMN2015}. However the RTT measured in our test-bed only includes eNodeB processing time and propagation time. Figure  \ref{fig:NGMN} illustrates the NGMN margin times for different PON fronthaul lengths, where the PON transmission latency of the different lengths are subtracted from the 500 us. The figure shows the average RTT for packet size 141 B for all three scenarios. The average RTT for the near ideal scenario exceeds the NGMN threshold at the fronthaul lenght approximately 17 km. For the cooperative DBA, the NGMN threshold is exceeded at 11 km. In the case of the un-synchronised scenario, the average RTT is never below the NGMN threshold.

The figures show as expected the best results in the case of the near ideal fronthaul. The lower latency is due to the LTE traffic being prioritised. This scenario will require a high degree of coordination to  align of the LTE and PON schedulers so that the upstream data arrives just as the PON frame is being packed. Also, if more LTE RUs/ONUs are associated with the same OLT, it might induce some delay anyway if more traffic has high priority.


\section{Conclusion}
\label{sec:conc}
The PON fronthaul is an economical way of using already established PON networks for fronthaul transmissions. This work demonstrates the use of XGS-PON for \ac{LTE} fronthaul by first proving the feasibility of transmitting Cloud-RAN MAC/PHY split over \ac{XGS-PON}, then it emulates a more realistic scenario where also background broadband traffic is included in the measurements.  \\ This work considers a cooperative DBA solution for upstream transmission from \ac{RU} to DU+CU where the latency is optimised in order to comply with mobile network latency requirements. The  test-bed used for the measurements consisted of an OAI setup connected by a virtual \ac{XGS-PON} network. The measurements showed satisfying results in terms of average RTT latency and average jitter. The initial feasibility studies without background traffic showed average RTTs below 150 us. From the results obtained including background traffic the maximum lenght of the XGS-PON fronthaul is unfeasible for the un-synchronised scenario, 11 km for the cooperative DBA scenario and 17 km for the near ideal scenario in order to stay below NGMN latency requirements.\\ This work proves the great potential for fronthaul transmission over PON and demonstrates how PON can be used to transport the MAC/PHY functional split. At the same time this work highlights the limitations in terms of latency when using PON as a fronthaul transport medium. We believe such number can be further increase with improvement in the synchronisation between DU and OLT, which will be carried out in future work.


 \section*{Acknowledgement}
Financial support from SFI grants 14/IA/252 (O'SHARE) and 13/RC/2077 and funding from BT and EPSRC iCASE award is gratefully acknowledged. 
\bibliographystyle{ieeetr}

\end{document}

%% file: acronym.tex
\begin{acronym}
\acro{AON}{Active Optical Network}
\acro{BBF}{BroadBand Forum}
\acro{BBU}{Base Band Unit}
\acro{BMap}{Bandwidth Map}
\acro{CO}{Central Office}
\acro{CORD}{Central Office Rearchitected as a Data Centre}
\acro{C-RAN}{Cloud Radio Access Network}
\acro{CU}{Centralised Unit}
\acro{DBA}{Dynamic Bandwidth Allocation}
\acro{DBRu}{Dynamic Bandwidth Report upstream}
\acro{DPDK}{Data Plane Development Kit}
\acro{DSL}{Digital Subscriber Line}
\acro{DU}{Distributed Unit}
\acro{EAL}{Environment Abstraction Layer}
\acro{FCC}{Federal Communications Commission}
\acro{FTTH}{Fiber-to-the-Home}
\acro{HARQ}{Hybrid Automatic Repeat Request}
\acro{KPI}{Key Performance Indicator}
\acro{InP}{Infrastructure Provider}
\acro{L2}{Layer-2}
\acro{L3}{Layer-3}
\acro{LTE}{Long Term Evolution}
\acro{MAC}{Medium Access Control}
\acro{NGA}{Next Generation Access}
\acro{ODN}{Optical Distribution Network}
\acro{OLO}{Other Licensed Operator}
\acro{OLT}{Optical Line Terminal}
\acro{ONU}{Optical Network Unit}
\acro{QoS}{Quality of Service}
\acro{PBMA}{Priority Based Merging Algorithm}
\acro{PON}{Passive Optical Network}
\acro{R-CORD}{Residential CORD}
\acro{RRU}{Remote Radio Unit}
\acro{RU}{Radio Unit}
\acro{SLA}{Service Level Agreement}
\acro{SRI-OV}{Single Root Input/Output Virtualization}
\acro{TDM}{Time Division Multiplexing}
\acro{T-CONT}{Transmission Container}
\acro{UE}{User Equipment}
\acro{URLLC}{Ultra Reliable Low Latency Communications}
\acro{vDBA}{virtual DBA}
\acro{VM}{Virtual Machine}
\acro{VNF}{Virtual Network Function}
\acro{VNO}{Virtual Network Operator}
\acro{vOLT}{Virtual OLT}
\acro{VULA}{Virtual Unbundled Local Access}
\acro{XG-PON}{10 Gigabit PON}
\acro{XGS-PON}{10 Gigabit per second Symmetrical PON}
\end{acronym}